\documentclass[twocolumn,showpacs,superscriptaddress,preprintnumbers,prb]{revtex4}
\usepackage{graphicx}
\usepackage{amsmath}

\begin{document}

\title{Microwave photoresistance of a high-mobility two-dimensional electron gas in a triangular antidot
lattice}
\author{Z. Q. Yuan}
\affiliation{Department of Physics and Astronomy, Rice University, Houston, Texas 77005,
USA}
\affiliation{Rice Quantum Institute, Rice University, Houston, Texas 77005, USA}
\author{C. L. Yang}
\affiliation{Department of Physics and Astronomy, Rice University, Houston, Texas 77005,
USA}
\affiliation{Rice Quantum Institute, Rice University, Houston, Texas 77005, USA}
\author{R. R. Du}
\affiliation{Department of Physics and Astronomy, Rice University, Houston, Texas 77005,
USA}
\affiliation{Rice Quantum Institute, Rice University, Houston, Texas 77005, USA}
\author{L. N. Pfeiffer}
\affiliation{Bell Laboratories, Lucent Technologies, Murray Hill, New Jersey 07974, USA}
\author{K. W. West}
\affiliation{Bell Laboratories, Lucent Technologies, Murray Hill, New Jersey 07974, USA}

\begin{abstract}
The microwave (MW)  photoresistance has been measured on a high-mobility 
two-dimensional electron gas patterned with a shallow triangular antidot lattice, 
where both the MW-induced resistance oscillations (MIRO) and magnetoplasmon (MP) 
resonance are observed superposing on sharp commensurate geometrical resonance (GR). 
Analysis shows that the MIRO, MP, and GR are decoupled from each 
other in these experiments.
\end{abstract}

\maketitle

The subject of electronic transport in a high-mobility two-dimensional electron gas
(2DEG) under microwave (MW) irradiation (frequency $\omega $) and a small
magnetic field ($B$) has attracted much recent attention, largely because of
the spectacular microwave-induced magneto-resistance oscillations \cite{1}(MIRO) 
and the subsequent zero-resistance states (ZRS) observed in very
clean samples.\cite{2,3,4,5,6,7} The period (in 1/$B$) of MIRO is controlled by a ratio
between the $\omega $ and the cyclotron frequency, $\epsilon = \omega /\omega_{c}$. To a large extent, present theoretical models proposed to
explain the MIRO are focused on the displacement of electrical currents by
random scatters,\cite{8,9} or on the electron distribution function modulated by the
presence of microwaves.\cite{10} These models are able to explain the observed
large amplitude oscillations, and suggest that in very clean samples and
under sufficient microwave power, the 2DEG can exhibit absolute negative
conductivity (ANC). The ZRS arise \cite{11} because in the ANC regime, the system
reaches an instability against the formation of current domains, and the
measured voltage in a macroscopic sample vanishes. Other models which do not
invoke ANC include those based on orbital dynamics \cite{12,13} or coherent interference effect.\cite{14} 

The MIRO and ZRS in a modulated 2DEG has been studied in recent theoretical work.\cite{15,16} 
In particular, for one-dimensional (1D) modulation,\cite{15} the displacement model\cite{8,9} and the distribution model\cite{10} are 
found to contribute to MIRO in an anisotropic manner. In principle, experiments in such a system 
can help to distinguish the relative contributions from each mechanism. For 2D modulation,\cite{16} different 
features can be expected on an ultraclean 2DEG. Moreover, a periodical  
modulation breaks translational symmetry of the 2DEG allowing studies of photoresistance with a 
finite momentum transfer $\Delta q=2\pi/a$ in this system. 

In this paper we report on an experimental study of MW photoresistance in a 2D 
periodically modulated 2DEG. The MIRO are observed in a triangular
antidotes array patterned in a high-mobility GaAs/Al$_{x}$Ga$_{1-x}$As heterojunction.
The MIRO retain a 1/$B$ periodicity as in unpatterned 2DEG, but their
amplitude is strongly damped. We also found the photoresistance peaks
corresponding to long-wavelength magnetoplasmon (MP) resonance; the dispersion of MP
is controlled by the width of the Hall bar rather than the period of potential
modulation. We conclude that the MIRO, MP, and DC geometric resonance (GR) are
decoupled from each other in these experiments. While MW photoconductivity has been reported for 
antidot arrays in a modest-mobility 2DEG,\cite{17} our work on a high-mobility system reveals different features, 
including MIRO, and opens a new window for studies of nonequilibrum quantum transport.

Our 2DEG is cleaved from a high-mobility GaAs/Al$_{0.3}$Ga$_{0.7}$As
heterostructure wafer grown by molecular beam epitaxy, with a $T = 0.3$ K mobility $\mu =1\times $10$^{7}$ cm$^{2}$/V s 
before processing. An 80-$\mu$m-wide, 320-$\mu$m-long Hall bar was first defined on the sample by optical lithography and wet
etching, then a triangular antidot lattice, with period $a = 1500$ nm and dot
diameter $d$ = 300 nm, was patterned on the Hall bar through $e$-beam
lithography and reactive ion etching [see inset (a) in Fig. 1]. After the
processing, the 2DEG has an electron density $n_e=2.83\times 10^{11}$ cm$^{-2}$
and a mobility $\mu =2.5\times 10^{6}$ cm$^{2}$/V s . Such parameters
were obtained after a brief illumination from a red light-emitting diode at $T$ = 4
K. Note that the introduction of antidot lattice reduced the mobility four
fold. However, the transport mean free path, $l=m^{\ast }v_{F}\mu
/e\sim$ 22 $\mu$m, where $m^{\ast}$ is the electron-effective mass and $v_F$ is 
the Fermi velocity, exceeds the period $a$ by at least one order of magnitude, indicating 
a much cleaner system than previously reported.\cite{17,18,19}

Our measurement was performed in a $^3$He refrigerator equipped with a
superconducting magnet. The MW was supplied by a set of Gunn diodes and sent 
via a waveguide to the sample immersed in the $^3$He coolant.
The mutual orientation of the waveguide, sample, and the magnetic field corresponds to
Faraday configuration; the excitation current flows perpendicularly to the
microwave polarization. The magnetoresisitance $R_{xx}$ was measured with a 
standard low-frequency lock-in technique (frequency 23 Hz and excitation current 1 $\mu$A).

To characterize the sample, at first we measured its magnetoresistance
without MW irradiation. A $R_{xx}$ trace is shown in Fig. 1, in which a set of
sharp peaks were resolved and are attributed to geometrical resonance. In magnetoresistance, 
GR occurs whenever the cyclotron radius $R_c$ becomes commensurate with the period \textit{a}
of the artificial scatterers,\cite{18,19} i.e.,
\begin{equation}
2R_{c}=\gamma a,
\end{equation}
where $\gamma $ is a constant depending on the geometry of the artificial
periodic scatterers, and $R_{c}=l_{B}^{2}k_{F}$ with $l_B=\sqrt{\hbar /eB}$ is the
magnetic length and $k_{F}=\sqrt{2\pi n_{e}}$ is the Fermi wave vector of the
2DEG. Note that $\omega _{c}\tau \geq 2\pi $ is satisfied in the magnetic-field regime 
of all these peaks, where $\tau $ is the momentum relaxation time determined by mobility, 
thus a full cyclotron orbit can be completed between the scattering events.

The observed GR (Fig. 1) are much sharper than previously reported\cite{18,19} and up to seven distinct 
resistance peaks are clearly resolved, attesting to the extraordinary quality of the sample. 
Remarkably, the GR peaks exhibit an alternating strength in a $B$ sweep, with the ``even'' 
peaks (i.e., 2,4,6) standing out as compared to the ``odd'' peaks (1,3,5,7).  Moreover, the peak position 
in $B$ does not conform to the set of ratio $R_c/a$ in previous reports.\cite{18,19}  
For example, according  to Ref. \onlinecite{19}, the strongest peak occurs at $R_c/a = 0.5$ corresponding to a first-pinned 
semiclassical orbit in a triangular lattice of period \textit{a}, whereas in Fig. 1 the strongest peak (labeled by 2) occurs at 
$R_c/a = 0.56$. A similar observation can be found for the second-strongest peak (4 in Fig. 1), where $R_c/a$ is 
found to shift from  $\sim$0.85 to  0.94 in the present case. The $\sim$10\% increase of the ratio $R_c/a$ indicates 
that in high-mobility samples, major resistance peaks may not correspond to pinned orbits. Rather, they correspond to 
those commensurate orbits, which dynamically experience most frequent scattering events, and hence higher conductance. 
Since in our high-mobility sample the GR occur in the regime $\rho_{xy} \gg \rho_{xx}$, and so, $\sigma_{xx} = \rho_{xx}/(\rho_{xx}^2+\rho_{xy}^2)$ 
$\propto$ $\rho_{xx}$, these states give rise to 
the high-resistance peaks in the Hall bar geometry. This scenario requires that the electron orbit of radius $R_c$ 
be scattered sequentially by multiple antidots. The orbital dynamics described here is equivalent to the theoretically studied runaway 
trajectories of delocalized electrons skipping from one antidot to another.\cite{19b, 19c} These orbits are drawn schematically 
in inset (b) of Fig. 1. The exceptional sharpness of the peak can be attributed to the small ratio of $d/a$ in the present sample. 

Our temperature-dependent measurements show that the major resistance peaks (2,4,6) persist up to above 10 K 
whereas the minor peaks (1,3,5) essentially diminish  at $T >$ 6 K. From these temperature dependences, we can roughly estimate the height 
of antidot potential for the electron orbits involved, which is much less than the Fermi energy of the 2DEG ($\sim$100 K).

\begin{figure}
\includegraphics{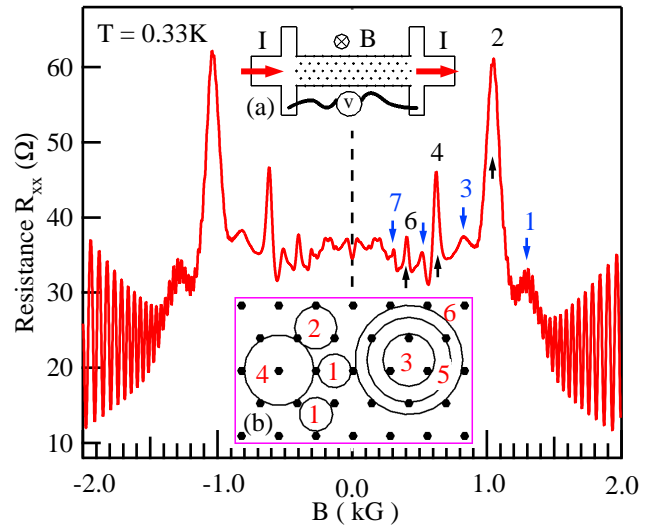}
\caption{\label{fig1} (Color online)Geometric resonance peaks in a low field of resistance $R_{xx}$ on 
2DEG with triangular antidote lattice (\textit{a} = 1500 nm,\textit{d} = 300 nm) at $T$ = 0.33 K. The peaks from 1 to 7 
correspond to ratio $R_c/a =$ 0.45, 0.56, 0.7, 0.94, 1.13, 1.45, and 1.9, respectively, 
where $R_c$ is the cyclotron radius. Inset (a) is the schematic experimental configuration of the sample 
and (b) is a sketch of commensurate orbits corresponding to the peaks.}
\end{figure}

Temperature dependence of the GR was measured from $T$ = 0.33 to 10 K; the results are shown 
in Fig. 2. At $T < 1.5$ K, the amplitude of the GR shows 
very little change with temperature, while the Shubnikov-de Haas oscillations dampen quickly with increasing 
temperature. At $T > 1.5$ K, the GR peaks are greatly damped by raising the temperature. This is clearly depicted 
in inset (c) of Fig. 2. This temperature dependence is consistent with the dominance of thermally excited phonon 
scattering at this temperature range,\cite{20}  rather than the mechanism of smearing of Fermi surface\cite{21} typically taking 
place at $T >$ 100 K. Interestingly, this behavior is similar to the temperature dependence of the $B$ = 0 
conductance [inset (b)], which is characteristic 
for the thermally excited phonon scattering model.\cite{20} 
From insets (a), (b), and (c), we can see that this regime transition happens at 
about $T = 1.5 \sim$ 2 K.
\begin{figure}
\includegraphics{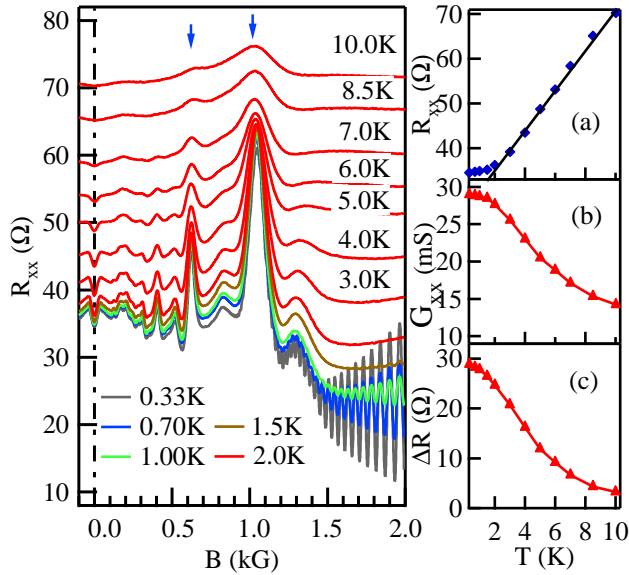}
\caption{\label{fig2}(Color online) Temperature dependence of geometric resonance for temperature 
$T$ = 0.33 to 10 K. Insets (a) and (b) are the resistance $R_{xx}$ and 
conductance $G_{xx}$ vs $T$, respectively, at $B$=0; 
inset (c) is the temperature dependence of the amplitude of GR peak 2 in Fig. 1.} 
\end{figure}

In the rest of the paper we shall report on photoresistance on the patterned 2DEG sample, 
where the photoresistance is defined as the change of resistance due to MW irradiation: 
$\Delta R_{xx}^{\omega}(B)=R_{xx}^{\omega }(B)-R_{xx}^{0}(B)$, $R_{xx}^{\omega}(B)$ and $R_{xx}^{0}(B)$ being 
the magnetoresistance with or without MW irradiation, respectively. 
As an example, Fig. 3(a) 
shows traces with and without continuous MW irradiation at $f$ = 56 GHz. 
In this measurement, the lattice temperature was kept at $T \sim 1$ K.

$R_{xx}$ trace with MW irradiation looks quite complicated because GR 
and MW-induced oscillations overlap in this low magnetic field regime. 
And as can be seen, the GR dominate this regime and make the MW-induced oscillation difficult 
to resolve. Our analysis shows that under MW irradiation, both the electron-heating effect and 
MIRO contribute to the $\Delta R_{xx}$ of the patterned  2DEG. The heating effect originates from 
the strong $T$ dependence of GR peaks in the temperature range $T >$ 1.5 K, as presented earlier. 
Since GR peaks exhibit a negative temperature coefficient, 
$dR_{xx}/dT < 0$, electron heating results in a reduction of $R_{xx}$ in the sample. 
Our main purpose in this experiment is to recover the photoresistance $\Delta R_{xx}^{\omega}$ from 
the background of the electron-heating signal.

It is possible to deduce the $\Delta R_{xx}^{\omega}$ by numerically substracting $R_{xx}^{0}$ from 
$R_{xx}^{\omega}$, as shown by the dotted line in Fig. 3(b). 
The $\Delta R_{xx}^{\omega}$ obtained in this way shows an oscillatory structure roughly periodical in 1/$B$,
similar to the MIOR observed in unpatterned  2DEG. On the other hand, significant distortion of the $R_{xx}^{\omega}$ 
can be anticipated owning to the contribution from the electron heating effect. While both the $R_{xx}^0$ and 
$R_{xx}^{\omega}$ data were taken at approximately the same lattice temperature $T \sim$ 1 K, the electron 
temperature is expected to be considerably higher in $R_{xx}^{\omega}$, leading to a large heating component. 
 
The $\Delta R_{xx}^{\omega}$ can also be measured directly by a double modulation technique\cite{22} in the following fashion. 
The MW was chopped at a frequency $f$ = 11.5 Hz, and the sample excitation current is synchronized at the 
double frequency 2$f$ = 23 Hz. Using the frequency $f$ as a lock-in reference, it can be shown that the $\Delta R_{xx}^{\omega}$ 
can be attained from the $90^{\circ}$ lock-in signal.\cite{22} 
The $\Delta R_{xx}^{\omega}$ at the same MW frequency 56 GHz, measured with a double-modulation method, is shown in 
Fig. 3(b). We note that the $\Delta R_{xx}^{\omega}$ obtained by both methods coincide reasonably well in their 
oscillatory structure in $B$, indicating the equivalence of these two methods in obtaining the $\Delta R_{xx}^{\omega}$. 

The following observations can be made for the MIRO in a modulated 2DEG, as measured here, in comparision to the MIRO 
in an unpatterned 2DEG.\cite{1} First, the MIRO in modulated 2DEG retain the characteristic period in $1/B$ 
determined by $\epsilon $ = $\omega $/$\omega_{c}$. Second, the damping of MIRO is much steeper in the modulated 
sample even though the mobility is comparable.  The mobility of the unpatterned sample in Ref. \onlinecite{1} is 
about $3\times10^{6}$ cm$^{2}$/V s, where up to ten MIRO peaks can be observed, whereas in the present sample of a 
mobility $2.5\times10^{6}$ cm$^{2}$/V s, only two peaks are clearly resolved. Of course, mobility is directly 
proportional to the transport lifetime $\tau_t$, which is only relevant at $B$ = 0. In a small magnetic field, 
the electron orbital dynamics are strongly influenced by the modulation. The above observations may point to an 
interesting theoretical question as to how such a modulation will affect the electrical current distribution in the MIRO 
regime. On a side note, our results could also support the notion that MIRO is a bulk, rather than an edge, effect. 
 
\begin{figure}
\includegraphics{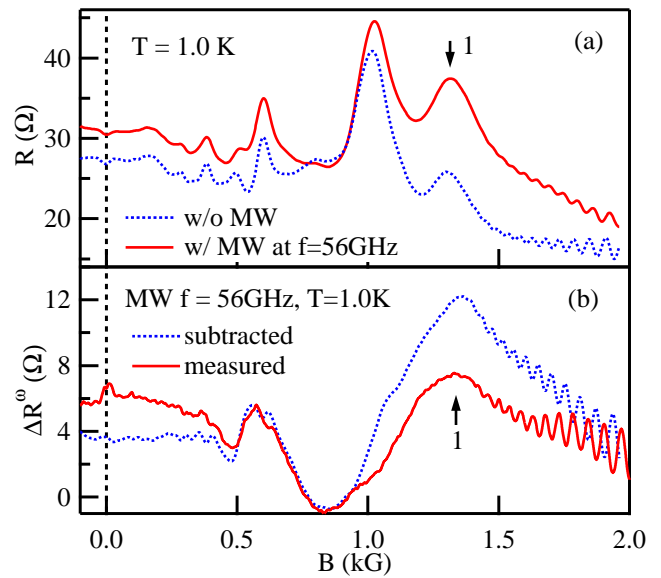}
\caption{\label{fig3}(Color online) (a) Magnetoresistance traces with and without continuous MW irradiation. 
(b) The difference (dotted line) between the two traces shown in (a), and the 
MW-induced signal (solid line) measured with the double-modulation technique. The arrows indicate the 
first peak of MW-induced oscillation.}
\end{figure}
\begin{figure}
\includegraphics{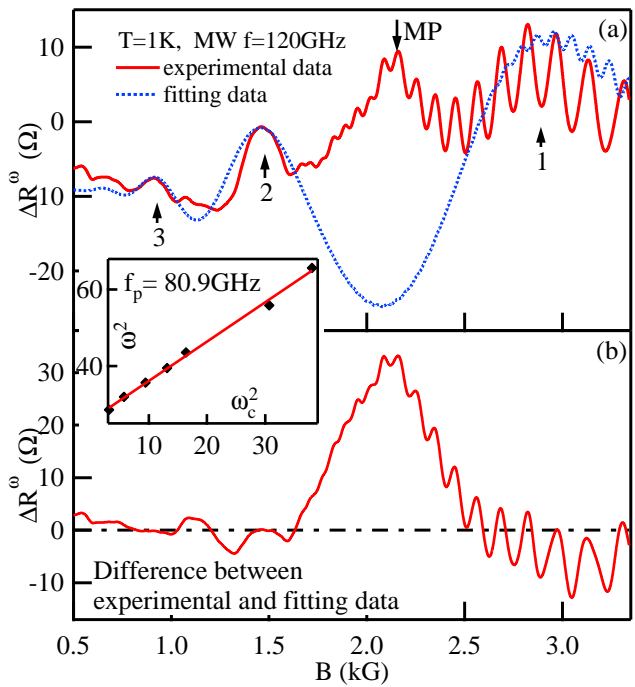}
\caption{\label{fig4} (Color online) A selected photoresistance trace at MW frequency $f$ = 120 GHz and the numerical fitting for 
MW-induced oscillations. 
The little arrows indicate the magnetoplasmon peaks. The difference between these two traces gives the clear 
magnetoplasmon peak [see (b)]. The inset shows the typical relation between the microwave frequency and magnetoplasmon 
peak position with the fitting curve based on eq. (2). The unit of axes is (100 GHz)$^2$.}
\end{figure}

Using the double-modulation method, we have measured the MIRO in a wide range of frequencies between 32 and 130 GHz, 
and qualitatively the same conclusion can be drawn throughout this range: MIRO can be clearly observed in a 2DEG 
with a shallow triangular antidots modulation, but with strongly damped amplitude. 

We observed an additional single resistance peak in the high-frequency regime, $f >$ 85 GHz, and identified it 
as the signal 
of magnetoplasmon resonance.\cite{1,23} Since the MIRO and MP resistances are superposed, we rely on the following 
empirical procedure to extract the MP signal. 

It is shown empirically in Ref. \onlinecite{24} that the periodicity and the shape of $\Delta R_{xx}^{\omega}$ 
in the MIRO can be reasonably fitted by a model based on the oscillatory electron distribution function at 
a given temperature $T$, with the Landau level width $\Gamma$ being a fitting parameter. In Fig. 4, we fit the 120 GHz 
experimental $\Delta R_{xx}^{\omega}$ (solid line) with a calculated curve (dotted line). Both the zero line and the 
amplitude of the curve were adjusted such that the envelops of both coincide. From the fit we obtain a width of 
$\Gamma$ $\sim$ 56 $\mu$eV $\sim$ 0.65 K, corresponding to a scattering time of $\hbar /\Gamma \sim$ 12 ps. It is 
interesting to compare the fitted $\Gamma$ in the MIRO of unpatterned 2DEG from the same wafer, which is 
41 $\mu$eV (16 ps). 

Subtracting the fitting curve from the experimental trace, we arrive at a trace which is dominated by a strong MP peak. 
For 2DEG, the magnetoplasmon resonance shall occur at
\begin{equation}
\omega ^{2}=\omega _{p}^{2}+\omega _{c}^{2},
\end{equation}
where $\omega $ is the MW frequency, $\omega _{c}=eB/m^{\ast }$ is the 
cyclotron frequency, and $\omega _{p}$ is the 2D plasmon frequency, which is given
by\cite{23}
\begin{equation}
\omega _{p}^{2}=n_{e}e^{2}k/2\epsilon \epsilon _{0}m^{\ast},
\end{equation}
where wave vector $k=\pi /w$ with \textit{w} the lateral width of 2DEG. The relation of Eq. (3) is well revealed 
in the inset of Fig. 4, and a fitting gives a 2D plasmon frequency of 80.9 GHz. For
our GaAs sample, taking into account the fact that the 2DEG is very close to the surface, the effective dielectric 
constant is 6.9.\cite{23} Using $m^{\ast}=0.068 m_{0}$ and the width of Hall bar is 80 $\mu$m , with formula (3) we
arrive at a 2D plasmon frequency 98 GHz, which is about 20\% larger than
the measured one. 

In summary, we have experimentally studied the microwave photoresistance in a high-mobility 2DEG modulated 
by a 2D triangular potential. The geometric resonance observed in the sample is remarkably different from 
previous systems of lower mobility. We observed microwave-induced resistance oscillations 
and magnetoplasmon resonance that are characteristically similar to those of unpatterned 2DEG. Present data shows 
that MIRO, MP, and geometric resonance are essentially decoupled from each other in these experiments. Ultimately, 
one would like to pursue the experimental regime where periodical modulation would lead to characteristically new 
behavior in the MIRO and ZRS. Along this line, a clean 2D electron system consisting of a short modulation period 
approaching magnetic length would offer exciting opportunities. Finally, the issue concerning scattering parameters 
and their influences in the microwave photoresistance remains open for experimental as well as theoretical work.

We thank K. Stone for technical assistance in the experimental work and D. Wu, and N. Zheng for sample fabrication. This work 
was supported by contract No. NSF-DMR 0408671.

\end{document}